\documentclass{article}

\usepackage{algorithm}
\usepackage{algpseudocode}
\usepackage{PRIMEarxiv}

\usepackage{algpseudocode}
\usepackage[utf8]{inputenc} % allow utf-8 input
\usepackage[T1]{fontenc}    % use 8-bit T1 fonts
\usepackage{hyperref}       % hyperlinks
\usepackage{url}            % simple URL typesetting
\usepackage{booktabs}       % professional-quality tables
\usepackage{amsfonts}       % blackboard math symbols
\usepackage{nicefrac}       % compact symbols for 1/2, etc.
\usepackage{microtype}      % microtypography
\usepackage{lipsum}
\usepackage{amsmath, amssymb, amsthm}
\usepackage{hyperref}
\usepackage{fancyhdr}       % header
\usepackage{graphicx}       % graphics
\graphicspath{{media/}}     % organize your images and other figures under media/ folder

\title{Analyzing Residential Speeding Using Connected Vehicle Data: A Case Study in Charlottesville, VA Area}

\author{
Shi Feng \\
    Department of Civil and Environmental Engineering\\
  University of Virginia, Charlottesville, VA 22903\\
  \texttt{sf5jh@virginia.edu} \\
   \And
 B. Brian Park, Ph.D. \\
  Link Lab\\
    Department of Civil and Environmental Engineering\\
    Department of Systems and Information Engineering\\
  University of Virginia, Charlottesville, VA 22903\\
  \texttt{bp6v@virginia.edu} \\
  \And
 Andrew Mondschein, Ph.D. \\
  Department of Urban and Environmental Planning\\
  University of Virginia, Charlottesville, VA 22903\\
  \texttt{mondschein@virginia.edu} \\
}

\begin{document}
\pagestyle{plain}
\maketitle
\begin{abstract}
This study uses connected vehicle data to analyze speeding behavior on residential roads. A scalable pipeline processes trajectory data and supplements missing speed limits to generate summaries at OpenStreetMap's way ID level. The findings reveal a highly skewed distribution of both aggressive and reckless speeding. Based on a case study of Charlottesville, VA's connected vehicle data on residential roads, we found that 38\% of segments had at least one instance of aggressive speeding, and 20\% had at least one instance of reckless speeding. In addition, night time speeding is 27 times more prevalent than day time, and extreme violations on specific road segments highlight how severe the issue can be. Several segments rank among the top 10 for both aggressive and reckless speedings, indicating that there exist high-risk residential roads. These findings support the need for both spatial and behavioral interventions. The analysis provides a rich foundation for policy and planning, offering a valuable complement to traditional enforcement and planning tools. In conclusion, this framework sets the foundation for future applications in traffic safety analytics, demonstrating the growing potential of telematics data to inform safer, more livable communities.
\end{abstract}

% keywords can be removed

\section{Introduction}
Residential speeding has been a persistent issue to the safety and livability. This problem is particularly pronounced in communities, where residential streets are frequently shared by vehicles, pedestrians, cyclists, and other vulnerable road user. Despite posted speed limits intended to promote safe travel, drivers often exceed these limits, increasing the risk of crashes, injuries, and fatalities \cite{nhtsa_speeding}.

Traditional methods for monitoring speeding behavior—such as police citations, static speed detectors, and crash reports—have significant limitations when applied to residential streets. Police citations capture only the subset of violators. Crash records, while valuable, represent only a fraction of safety-critical events, as they are documented only after incidents involving sufficient property damage or injury. Moreover, these records are inherently retrospective and do not capture the actual driving behavior such as speeding. Similarly, speed cameras are constrained, providing data for a limited number of locations and failing to capture broader driving patterns within the road network, especially in residential roads.

Moreover, large-scale vehicle trajectory datasets commonly used in transportation research are often designed for freeway or arterial analysis and lack sufficient granularity on lower-speed residential roadways. These datasets typically under-represent the fine-scale, everyday driving behaviors occurring on neighborhood streets.

Plus, The Highway Capacity Manual (HCM) \cite{HCM} and the Highway Safety Manual (HSM) \cite{HSM}, two foundational references in transportation engineering, do not include dedicated chapters addressing residential roads. As a result, critical operational and safety characteristics specific to these roadways—such as localized speeding behavior—are not systematically studied within those frameworks. Given the importance of residential streets for pedestrian safety, neighborhood livability, and traffic calming, there is a clear need for additional analysis to understand and address speeding behavior on these roads.

%new
Several national and international guidance documents, however, address speed management in residential environments. FHWA’s Speed Management Countermeasures identifies traffic-calming treatments, such as speed humps, that are commonly applied on residential streets and have been shown to reduce operating speeds by nearly 10mph \cite{FHWA_SpeedCountermeasures}. The Speed Management: A Road Safety Manual for Decision-Makers and Practitioners emphasizes the vulnerability of pedestrians and cyclists on residential streets and recommends much lower maximum speeds in residential areas with frequent pedestrian activity \cite{GRSP_2008}. In addition, the California Manual for Setting Speed Limits (2020) and the NCHRP Posted Speed Limit Setting Procedure and Tool (2021) incorporate roadway context, access density, roadside activity, and pedestrian presence when establishing posted limits on urban and suburban local streets \cite{CA_SpeedManual2020, NCHRP_966}. These documents primarily provide detailed engineering and policy guidance for setting or managing speeds, but they do not offer trajectory-level evidence on actual speeding behavior. 
%new ends

In addition, widely used traffic information systems, such as Google Maps, or most navigation platforms, are designed to present congestion and travel times—not speeding behavior. These systems do not provide insight into whether drivers are complying with posted speed limits. As a result, the problem of residential speeding remains largely invisible in the tools commonly used by the public and transportation agencies.

To further investigate speeding concerns, we examined the public crash data from the Virginia Department of Transportation (VDOT) in Charlottesville and Albemarle County for the year of 2022 \cite{crashdata}. However, only 984 unique way IDs (out of 41,124 way IDs in total) were associated with reported crashes, and just 172 of these were identified as residential road segments in OpenStreetMap (out of 4,677 residential way IDs in total). As illustrated in Table \ref{tab:crash_residential_wayids}, the majority of the residential way IDs do not have crash record information in 2022. For the residential way IDs with crash records, the majority of these residential segments experienced only a single or two crashes, with very few segments reporting more than two crashes. This limited coverage and sparse distribution make it nearly impossible to identify meaningful patterns in speeding behavior or assess speeding severity using crash data alone. Relying solely on crash data from residential roads would likely overlook many locations where dangerous speeding is prevalent but has not yet resulted in a reportable incident, as not all way IDs have reported crash records, where 40,140 way IDs lack crash information, including 4,645 residential way IDs with no recorded crashes.

% \begin{figure}[!ht]
%     \centering
%     \includegraphics[width=0.8\textwidth]{crash_bar.png}
%     \caption{Distribution of crash frequency on residential road segments}
%     \label{fig:res_crash_bar}
% \end{figure}

% \begin{table}[ht]
% \centering
% \caption{Crash Frequency Distribution on All Way IDs in Charlottesville and Albemarle County in 2022}
% \begin{tabular}{|c|c|}
% \hline
% \textbf{\# of Crashes} & \textbf{\# of Way IDs} \\
% \hline
% 0   &   40,140\\
% 1 & 536\\
% 2 & 186\\
% 3 & 77\\
% 4 & 46\\
% 5 & 36\\
% 6-10   &  63 \\
% 11-20   &   28\\
% 21-30   &   7\\
% 31-40   &   5\\

% \hline
% \textbf{Total} &   41,124\\
% \hline
% \end{tabular}
% \label{tab:crash_all_wayids}
% \end{table}

\begin{table}[ht]
\centering
\caption{Crash Frequency Distribution on Residential Way IDs in Charlottesville and Albemarle County in 2022}
\begin{tabular}{|c|c|}
\hline
\textbf{\# of Crashes} & \textbf{\# of Residential Way IDs} \\
\hline
1   &   140\\
2   &   28\\
3   &   1\\
4   &   1\\
5   &   2\\
No Info   &   4,645\\

\hline
\end{tabular}
\label{tab:crash_residential_wayids}
\end{table}

The emergence of connected vehicle telematics data offers a transformative opportunity to overcome these limitations. Telematics data captures timestamped, high-frequency vehicle trajectories at a network-wide scale, covering both major and minor roads. Unlike traditional data sources, this information provides detailed insights into real-world driving behaviors, including instantaneous speed relative to posted limits. When combined with open-source mapping platforms like OpenStreetMap (OSM) \cite{openstreetmap}, telematics data can be leveraged to monitor speeding behavior continuously and comprehensively across residential street networks.

%\subsection{Objectives}
% **I like this. Please consider one of the objectives is to investigate speeding behaviors (frequencies and seriousness) **

% Here I will write about the goals of this paper: using Wejo telematics data to identify, quantify, and visualize residential speeding behavior.

% Detect instances of speeding and extreme speeding across a large volume of connected vehicle data provided by Wejo.

% Focus the analysis on residential roads using an example of Charlottesville and Albemarle County, defined via OpenStreetMap metadata (speed limits <= 40 km/h or missing).

% Develop visualization of speeding patterns to show the severity of speeding issues.

This study aims to leverage connected vehicle data to identify, quantify, and visualize residential speeding behavior within Charlottesville and Albemarle County. This area was chosen as a case study because it offers a wide variety of residential street types—ranging from dense neighborhoods to more spread-out suburban roads—within a relatively small region, making it practical for focused analysis. Specifically, the objectives of this research are as follows:

\begin{itemize}

\item Detection of Speeding Behavior: Identify occurrences of aggressive speeding (exceeding 10 mph than the speed limit) and reckless speeding (exceeding 20 mph than the speed limit) based on a large sample of connected vehicle trajectories.

\item Focus on Residential Roads: Limit the analysis to roadways classified as residential, using information derived from OSM.

\item Investigation of Speeding Behavior Frequency and Severity: Examine how frequently residential speeding occurs and assess the seriousness of these behaviors. This includes analyzing the distribution of speeding rates across different road segments.

% \item Visualization of Speeding Patterns: Develop an intuitive, map-based visualization approach that highlights the spatial distribution and severity of speeding behaviors at the way ID and Vehicle ID level, allowing stakeholders to easily interpret where speeding is most prevalent within the residential network.
\end{itemize}
%\subsection{Contributions}
% Here I will write about the key methodological and practical contributions of this research to the transportation analytics and safety community.

% A practical framework for cleaning, enriching, and analyzing Wejo vehicle data for residential roads.

% A OSM based mapping approach to visualize speeding behavior at the way ID level.

% Empirical evidence of speeding patterns that can inform local interventions and policy.

This research makes several important methodological and practical contributions to the transportation safety and analytics community:
\begin{itemize}
    \item A Scalable Analytical Framework: The study proposes a reproducible pipeline for cleaning, enriching, and analyzing connected vehicle data specifically targeted at residential road networks. 

   \item Cumulative Distribution Function Visualization: By integrating telematics data with OSM-based mapping, the study introduces a practical visualization method for depicting speeding rates across individual road segments (way IDs) rather than aggregated at larger geographic levels, enabling fine-grained spatial analysis.

  \item Empirical Insights for Policy and Planning: The analysis provides data-driven evidence of speeding patterns on residential streets in Charlottesville and Albemarle County. These findings have direct implications for local agencies, supporting targeted interventions such as speed enforcement, traffic calming measures, or policy development aimed at improving neighborhood safety.

\end{itemize}

\section{Literature Review}
%(First draft done)
Research on speeding issues by analyzing vehicle data has covered a few overlapping areas, contributing to the understanding of how and why drivers speed.

By analyzing spatial patterns and real-world telematics data, studies have uncovered clear differences in speeding intensity from one neighborhood to another. Alrassy et al. analyzed nearly 4,000 New York City fleet vehicles’ telematics—extracting speed variation and hard-braking and acceleration rate—and demonstrated that there are moderate correlations between the safety surrogate measures and collision rates \cite{Alrassy2023Driver}. Xiang et al. developed a GeoSpatial and Temporal Mapping of Urban Mobility  framework, which converts trajectory data into driving metrics including time of idling, cruising, acceleration, and deceleration; their results illustrate strong correlation between vehicle road occupancy and aggressive driving \cite{Xiang2024Mapping}. Cai et al.  integrated second-by-second driving speed data from hundreds of thousands of drivers across ten U.S. cities with enforcement records, showing that current enforcement practices across race and ethnicity \cite{Cai2022Measuring}. While these studies demonstrate the value of telematics for uncovering neighborhood-level driving risk, most have focused on commercial fleets or dense urban cores; our study builds on this foundation by turning attention to residential roads, where infrastructure and driver behavior may differ significantly.

Driver-centric data driven models and analysis also help explain why certain drivers accelerate or exceed speed limits. Moosavi and Ramnath proposes a scalable framework for predicting driving risk using telematics and contextual data to represent driving styles as tensors, which are then clustered into risk cohorts based on behavioral similarity. The model enables accurate, driver-specific risk assessment useful for insurance pricing and self-improvement \cite{Moosavi2023Context}. Pérez-Marín et al. used quantile regression on real telematics data from usage-based insurance drivers to analyze how distance driven above the speed limit varies by total distance, urban/nighttime driving, and gender. The findings show that speeding risk is heterogeneous, with particularly high-risk behaviors concentrated among male drivers in non-urban and nighttime contexts—suggesting that targeted safety interventions are needed \cite{PerezMarin2019Quantile}. Our work complements this perspective by aggregating and analyzing observed speeding behavior across physical road segments, offering insights not only into who speeds, but also where speeding problems are concentrated within the residential road network.

Trip-level studies, using techniques such as motif detection, surveys, and controlled experiments, help pinpoint specific driving maneuvers and situational factors that are closely related to speeding behavior. Silva and Henriques used a novel approach to identifying driving maneuvers from telematics data using time-series motif detection, specifically a modified Extended Motif Discovery (EMD) algorithm. Applied to a naturalistic driving dataset, the method detected both basic and complex maneuvers, demonstrating the potential of motif discovery for deeper behavioral analysis in road safety research.\cite{Silva2020Finding}. Kontaxi et al. leveraged high‐resolution smartphone data from 88 drivers, along with self‐reported questionnaires, to fit generalized linear mixed‐effects models of the percentage of time spent over speed limits. Their results show that travel distance, use of the mobile phone while driving, male sex, and younger age (18 to 34) are positively associated with higher proportions of speeding time, while drivers who report 'never or rarely' speeding exhibit significantly lower speeding shares \cite{Kontaxi2023Exploring}. Lauridsen et al. conducted a controlled field experiment in Copenhagen, pairing smartphone telematics with auditory warning interventions. They showed that alert design and driver experience interact to influence the frequency and duration of speeding on road segments \cite{Lauridsen2024Data}. Unlike studies that analyze specific trip segments or maneuvers, our approach generalizes speeding behavior across space and time by summarizing trends across thousands of connected vehicle records, with emphasis on identifying speeding issues on residential roads.

Prior studies have also examined speeding behavior on residential roads through behavioral and policy-focused lenses. Dinh and Kubota applied the theory of planned behavior to assess why drivers willingly exceed 30 km/h limits on Japanese residential streets, finding that perceived appropriateness of the speed limit, perceived function of residential streets, and perceived right of vulnerable street users significantly influence both speeding intention and observed behavior \cite{DINH2013199}. Meanwhile, Islam et al. evaluated the impact of lowering posted speed limits from 50 to 40 km/h in residential neighborhoods and found statistically significant reductions in mean speed and speed variance, even though full compliance remained low \cite{ISLAM2014483}. These findings underscore the importance of aligning both attitudes and policy interventions when addressing residential speeding—a theme observed in our analysis—from the planning side.

Despite these advances, few studies directly quantify the frequency, duration, and spatial distribution of speeding on residential roads in suburban areas—and even fewer combine real world vehicle telematics data analysis. Our work leverages trajectory dataset to capture frequency and severity of speed limit violation in residential segments in suburban areas of the United States. By finding trajectory-derived speeding issues and generalizing findings within Albemarle County and Charlottesville City in suburban Virginia, we address critical gaps in the literature on residential‐road speeding issues and establish a foundation for data-driven safety interventions.

\section{Methodology}
% \subsection{Data Description}
% Here I will write about the Wejo data format, what attributes are available, and how they are used in this analysis.

% Trajectory-level vehicle records with timestamp every 5 sec, location, speed, and matched way ID.

% Way ID and osm speed limit are used to contextualize speeding behavior.

This analysis relies on high-frequency connected vehicle trajectory data. The dataset contains trajectory-level records, with each observation capturing a vehicle’s state collected every 3 seconds. 14-day vehicle trajectory data from April 2022 was analyzed, and each record includes essential attributes such as:
\begin{itemize}
    \item Timestamp (with date and time in 1s precision)
    \item Geographic coordinates (latitude and longitude)
    \item Instantaneous speed (in km/h)
    \item Way ID (a unique identifier linked to OSM road segments)
    \item Postal Code
\end{itemize}

These attributes are critical for contextualizing speeding behavior, particularly for assessing speeding relative to posted limits. 

% \subsection{Speed Limit Augmentation}
% Here I will write about the process of enriching the dataset with an added speed limit column.

% If osm speed limit is present, use it directly. If missing, default to 25 mph (40 km/h) (assumed residential).

% (Justification for using 25 mph (40 km/h) as a conservative baseline if needed)

The presence of Way ID in the dataset enables spatial association with roadway geometry, providing a direct link to OSM attributes, including speed limits and road classifications. The speed limit information, when available, provides the regulatory baseline against which speeding behavior is evaluated.

To ensure the accuracy and relevance of the speeding analysis, the dataset undergoes a multi-step filtering and augmentation process:

\begin{itemize}
    \item Geographic Filtering by Postal Code: the first step restricts the dataset to vehicle trajectories occurring within the Charlottesville and Albemarle County area. This is achieved by filtering the dataset based on the postal codes associated with each data point, ensuring that only trips within the intended geographic boundaries are retained.

 \item Road Type Filtering Using Way ID: After the geographic filter, the dataset is further refined to include only trips occurring on residential roadways. This is accomplished by cross-referencing each Way ID with OSM road type classifications. Only those way IDs tagged as "residential" in OSM are retained for the analysis. This step ensures that higher-speed facilities, such as arterials or highways, are excluded from the residential speeding analysis.

 \item Speed Limit Augmentation for Way IDs: Once the dataset is spatially and contextually filtered to include only residential roads within Charlottesville and Albemarle County, the next step addresses gaps in the OSM speed limit data. Not all way IDs within the filtered dataset have an associated posted speed limit recorded in OSM. To account for these missing values, the following augmentation rule is applied:
\begin{itemize}
    \item  If the OSM speed limit is present in the maxspeed tag, it is directly adopted as the added speed limit for that road segment.
    \item If the speed limit is missing, a default value of 25 miles per hour (40 kilometers per hour) is assigned. This assumption prioritizes safety and aligns with the statutory residential speed limit in Virginia \cite{vdot_speed_limits}, including the jurisdictions of Charlottesville and Albemarle County.
\end{itemize}
\end{itemize}

The resulting added speed limit column ensures that every road segment in the dataset has an assigned speed limit, enabling a consistent and comprehensive evaluation of speeding behavior across all residential roads within the study area.

% \subsection{Speeding Behavior Identification}
% Here I will write about the criteria for classifying vehicle behavior into threshold and extreme speeding.

% Threshold speeding: speed >= added speed limit +15mph.

% Extreme speeding: speed > 2 × added speed limit.

In this research, speeding behavior is classified using two distinct thresholds to capture both moderate and severe violations of speed limits:

\begin{itemize}
    \item Reckless Speeding: Defined as instances where a vehicle’s instantaneous speed is greater than or equal to the posted speed limit plus 20 miles per hour, this threshold aligns with the Commonwealth of Virginia’s statutory definition of reckless driving \cite{virginia_code_862}.
    
    \item {Aggressive Speeding: While reckless speeding is commonly defined as exceeding the speed limit by 20 mph or more, aggressive speeding in this study is defined as exceeding the posted speed limit by 10 mph or more.}

\end{itemize}

The aggressive and reckless speeding categories are not mutually exclusive. Additionally, the posted speed limits originally in mph are converted to km/h prior to comparison with the connected vehicle speeds, which are reported in km/h.

% \subsection{Aggregation of Speeding Statistics}
% Here I will write about how the data is aggregated at the way ID level and vehicle ID level.

% Count the total number of rows (observations) per way ID and vehicle ID traveling on residential roads.

% Calculate the percentage of rows exceeding threshold and extreme speeding criteria.

% Create a summary spreadsheet with way ID, vehicle ID, and meaning of each column.

After identifying speeding behavior at the trajectory point level, the analysis proceeds to aggregate the results into summary output at the way ID level. This approach allows the analysis to capture spatial patterns of speeding across the road network.

The Way ID--based output file summarizes speeding behavior aggregated for each unique OSM Way ID, which corresponds to a distinct road segment. This file includes the following metrics:

\begin{itemize}
    \item \texttt{OSM\_way\_id}: Unique identifier for the road segment from OSM.
    \item \texttt{total\_row\_number}: Total number of trajectory points  recorded on this way ID.

    \item \texttt{added\_speed\_limit}: Speed limit applied to this way ID, either sourced directly from OSM or imputed as 25 mph if missing.
    
    \item \texttt{total\_morning\_count}: Total number of trajectory points  recorded on this way ID during daytime hours (8 AM to 4 PM).
    
    \item \texttt{total\_night\_count}: Total number of trajectory points  recorded on this way ID during nighttime hours (9 PM to 5 AM).
    
    \item \texttt{morning\_speeding\_count}: Number of aggressive speeding points occurring during daytime hours (8 AM to 4 PM).
    \item \texttt{night\_speeding\_count}: Number of aggressive speeding points occurring during nighttime hours (9 PM to 5 AM).
    
    \item \texttt{way\_id\_avg\_speed}: Average vehicle speed (km/h) observed on this way ID.
    
    \item \texttt{way\_id\_speed\_sd}: Standard deviation of vehicle speeds on this way ID, reflecting variability in speed behavior.

    \item \texttt{aggressive\_speeding\_row\_number}: Number of trajectory points where vehicles exceeded the aggressive speeding definition (speed $\geq$ added speed limit + 10 mph).
    
    \item \texttt{reckless\_speeding\_row\_number}: Number of trajectory points where vehicles exceeded the reckless speeding definition (speed $\geq$ added speed limit + 20 mph).

    % \item \texttt{way\_id\_length}: Segment length of this way ID.

    \item \texttt{aggressive\_speeding\_percent}: Percentage of trajectory points engaged in the aggressive speeding. Defined as:
    \[
    \frac{\texttt{aggressive\_speeding\_row\_number}}{\texttt{total\_row\_number}} \times 100\%
    \].
    
    \item \texttt{reckless\_speeding\_percent}: Percentage of trajectory points engaged in reckless speeding. Defined as:
    \[
    \frac{\texttt{reckless\_speeding\_row\_number}}{\texttt{total\_row\_number}} \times 100\%
    \].

\end{itemize}

This output provides a comprehensive view of how frequently speeding occurs on individual residential road segments, how it varies by time of day, and the overall speed distribution characteristics.

\section{Results}

% To ensure sufficient trajectory representation for each segment, we first examined the distribution of the total number of data points (\texttt{total\_row\_number}) across all 5,061 residential way IDs. As shown in Figure \ref{fig:wayid_row_percentage}, the majority of segments have a substantial number of data points, but a nontrivial share (10th percentile) have fewer than 16 observations.  As such, we excluded these way IDs falling below the 10th percentile of total row count (16 rows), reducing the sample to 4,557 segments for the aggressive speeding analysis.
% \begin{figure}[!ht]
%   \centering
%   \includegraphics[width=0.8\textwidth]{wayid_row_percentage.png}
%   \caption{Percentage Distribution of Way IDs by Total Row Count} 
%   \label{fig:wayid_row_percentage}
% \end{figure}
\subsection{Aggressive Speeding}
To ensure sufficient trajectory representation for each segment, we examined the distribution of total data points (\texttt{total\_row\_number}) across all 5,061 residential way IDs. While most segments were well-covered, a substantial minority had relatively sparse coverage. To maintain analytical rigor and avoid inflated percentage due to inadequate sampling, we excluded segments with fewer than 100 observations. This threshold roughly corresponds to the lower 25th percentile of total row counts and ensures at least moderate coverage over a 14-day period—especially considering the average segment length of approximately 860 meters. Applying this filter reduced the sample to 3,754 segments for the aggressive and reckless speeding analysis. Among the 3,754 residential segments retained for analysis, approximately 97.2\% had missing speed limit information in OSM and therefore received the 25 mph imputation.

\begin{figure}[!ht]
  \centering
  \includegraphics[width=0.8\textwidth]{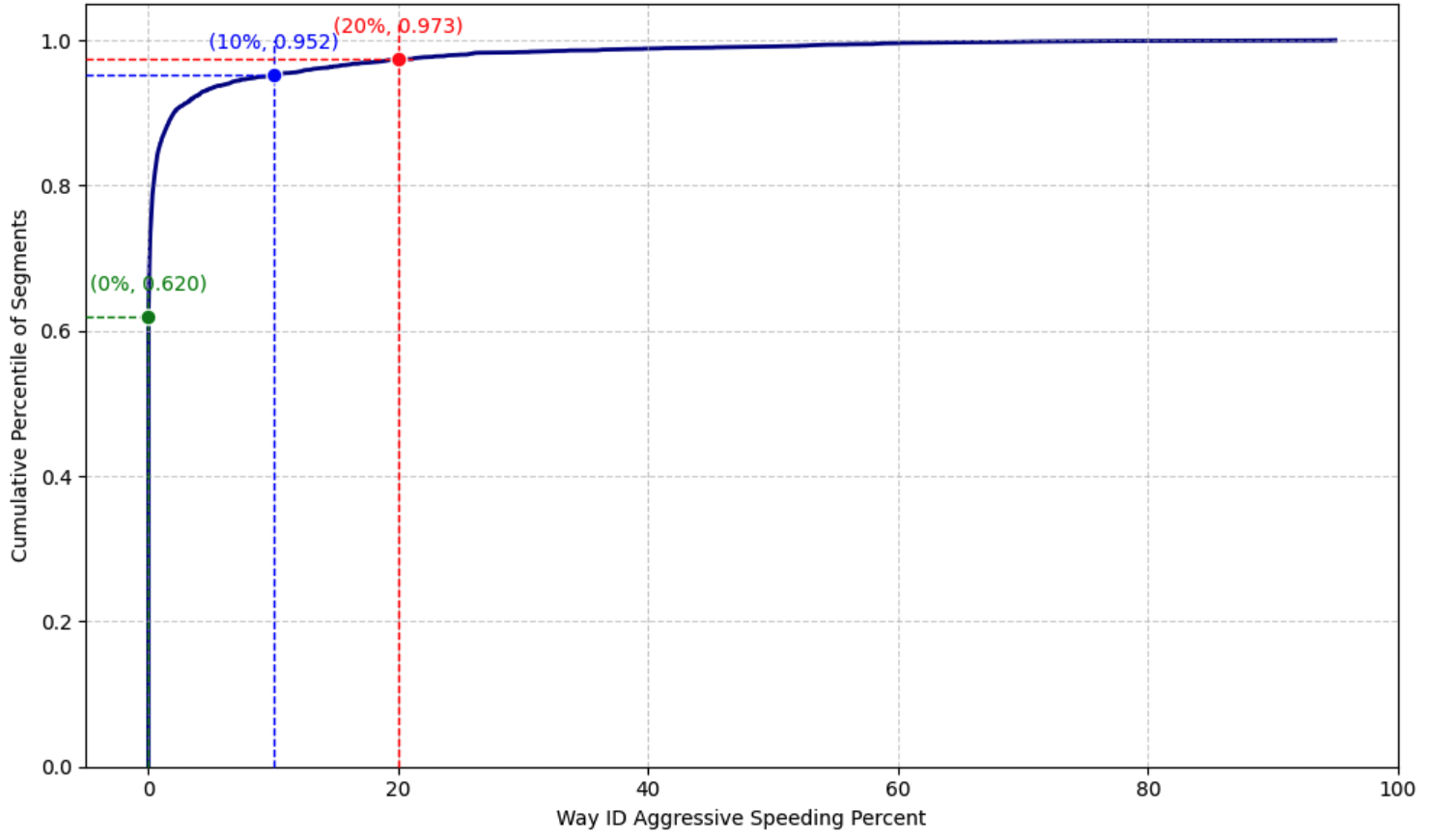}
  \caption{Cumulative Distribution of the aggressive speeding \% for each Way ID} 
  \label{fig:cdf_wayid_aggressive}
\end{figure}

 To explore the severity and prevalence of aggressive speeding behavior across residential roads, we also analyzed the cumulative distribution of speeding percentages for each road segment. The x-axis represents the percentage of recorded trajectory points that exceeded the aggressive speeding threshold on each segment, while the y-axis shows the cumulative percentile of segments. As shown in Figure \ref{fig:cdf_wayid_aggressive},  the distribution is also heavily right-skewed: 38\% of segments have at least one aggressive speeding point. Additionally, 4.8\% of segments recorded aggressive speeding on at least 10\% of their trajectory points, while in more extreme cases, 2.7\% of segments exceeded 20\%. This steep tail highlights how a fraction of residential roads accounts for a disproportionate share of observed speeding. 
 % The cumulative distribution offers a more continuous view of risk distribution and reinforces the importance of prioritizing a small number of high-violation segments for targeted safety interventions.

Rather than presenting a full set of summary statistics—which may not be meaningful given the large number of road segments—this analysis focuses on highlighting the most critical cases. Table~\ref{tab:top_wayid} lists the top 10 residential road segments with the highest aggressive speeding percentage, alongside key attributes such as total row counts, total vehicles, average and standard deviation of vehicle speeds (including all recorded speeds, whether they exceed the aggressive speeding threshold or not). To ensure privacy, these way IDs are replaced with letters.

\begin{table}[!ht]
	\caption{Top 10 Residential Road Segments with the Highest Aggressive \% for each Way ID}
	\label{tab:top_wayid}
	\begin{center}
    \small
		\begin{tabular}{l l l l l l l} \hline
			Way ID & Total Rows  & Aggressive\% & Avg Speed (km/h) & Speed SD (km/h) & {Speed Limit (km/h)}\\\hline
			a   & 100                & 95\%             & 79.11     & 18.28  &{40}  \\
			b   & 573             & 94.24\%             & 69.73     & 8.98   &{40} \\
			c   & 335               & 90.15\%             & 70.19     & 17.12 & {40}   \\
			d  & 274             & 85.04\%             & 72.17     & 19.73   &{40} \\
			e  & 7495                & 77.41\%             & 64.41     & 21.84 &{40}   \\
            f & 5295                & 74.69\%             & 62.4      & 24.57 &{40}\\
            g & 3312                   & 73.67\%               & 62.29     & 23.49 &{40}\\
            h & 36342                  & 71.04\%           & 56.94    & 26.52 &{40}\\
            i & 1223                  & 69.91\%             & 59.66    & 24.13 &{40}\\
            j & 8433                   & 69.04\%            & 59.39    & 20.11 &{40}\\
            \hline
		\end{tabular}
	\end{center}
\end{table}

\subsection{Reckless Speeding}
\begin{figure}[!ht]
  \centering
  \includegraphics[width=0.8\textwidth]{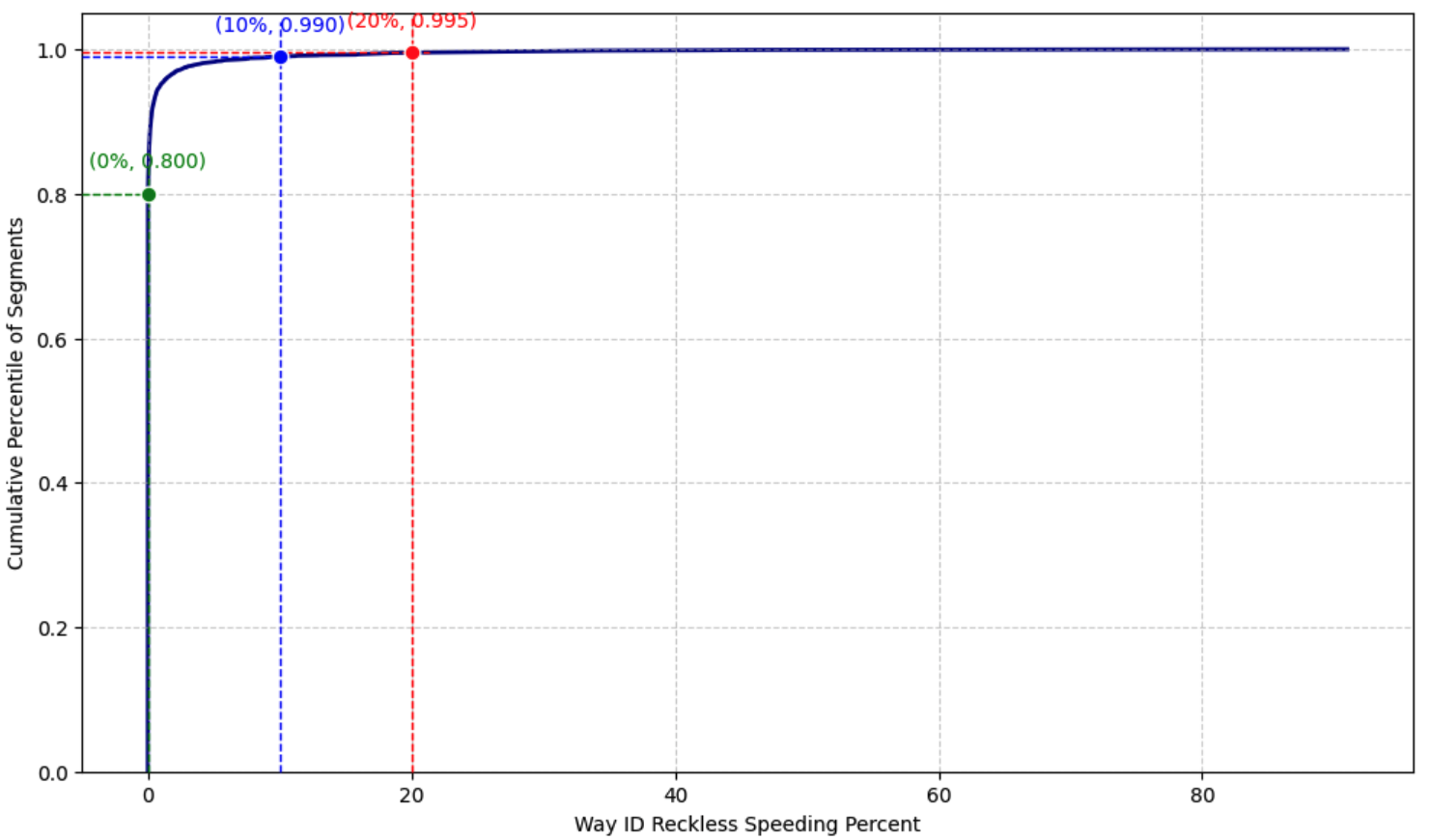}
  \caption{Cumulative Distribution of the reckless speeding \% for each Way ID} \label{fig:cdf_wayid_reckless}
\end{figure}

Similarly, Figure~\ref{fig:cdf_wayid_reckless} illustrates the cumulative distribution of reckless speeding percentage across residential road segments. The distribution is even more sharply right-skewed: 20\% of segments recorded at least one instance of reckless speeding. Notably, the top 1\% of segments (35 segments) exhibited reckless speeding percentage exceeding 10\%. This steep curve emphasizes the rarity of such extreme violations, but their severity—defined by substantial exceedance of the posted speed limit—underscores their significance as a critical safety concern.

% Figure \ref{fig:reckless_frequency_histogram} presents the histogram of reckless speeding observance percentage, defined as the proportion of trajectory points on each segment that exceed the posted speed limit by at least 20 mph. Like the aggressive speeding distribution, this frequency histogram (binned by 1\%) is heavily right-skewed. Out of the 4,557 retained way IDs, 562 segments (12.3\%) exhibit only 0–1\% reckless speeding observance. The frequency drops sharply beyond this point, with only sparse counts observed as the speeding percentage increases. A handful of road segments—most notably one showing 100\% reckless speeding—indicate consistent and pervasive violations at high speeds. While the tail is thinner than that of the aggressive speeding distribution, a moderate number of segments (10\%–30\%) still display noteworthy levels of reckless speeding, meriting further attention due to their potential safety implications.

Similarly, table ~\ref{tab:top_wayid_extreme} lists the top 10 residential road segments with the highest reckless speeding percentage. The average speed and standard deviation column also include all recorded speeds, whether they exceed the aggressive speeding threshold or not. {Way IDs a, b, c, d, e, f, g, h, and i represent the same segment in the aggressive speeding table ~\ref{tab:top_wayid}.}

\begin{table}[!ht]
	\caption{Top 10 Residential Road Segments with the Highest Reckless Speeding Percentage}
	\label{tab:top_wayid_extreme}
	\begin{center}
    \small
		\begin{tabular}{l l l l l l l}\hline
			Way ID & Total Rows  & Reckless\% & Avg Speed (km/h) & Speed SD (km/h) & Speed Limit (km/h)\\\hline
			a   & 100                & 91\%             & 79.11     & 18.28 &{40}   \\
            d   & 274               & 64.6\%             & 72.17     & 19.73  &{40}  \\
            c & 335                & 60\%             &70.19      &17.12 &{40} \\
            f & 5295                  &47.5\%               &62.4     &24.57 & {40}\\
			b   & 573             & 44.33\%             & 69.73     & 8.98   & {40} \\
            g  & 3312               &42.66\%              &62.29        &23.49 &{40}\\
			e  & 7495              & 39\%             & 64.41     & 21.84  &{40}  \\
			h  & 36342                & 33.98\%             & 56.94     & 26.52 &{40}  \\
            i  &1223                  &33.93\%        &59.66    &24.13 &{40}\\
            k &854               &31.15\%              &58.40      &25.56 &{40}\\
            \hline
		\end{tabular}
	\end{center}
\end{table}
\newpage

% \subsection{Spatial Map Output}
% To further analyze speeding patterns, an aggregated map was developed where each OSM way ID is represented by a colored marker. The color of each marker indicates the percentage of aggressive or reckless speeding observance occurring on that road segment. Darker colors represent a higher percentage of speeding events, while lighter colors indicate lower percentages.

% Additionally, each marker includes detailed mouse-hovering information, displaying the following:
% \begin{itemize}
% \item Way ID
% \item Total number of trajectory points on this way ID
% \item Aggressive (Reckless) speeding percentage
% \end{itemize}

% This map enables quick identification of roads with disproportionately high rates of speeding, providing valuable insights for targeted safety interventions or further analysis.

% A sample of the aggressive speeding and reckless speeding map is shown below:

% \begin{figure}[!ht]
%   \centering
%   \includegraphics[width=0.8\textwidth]{threshold_speeding_osm.png}
%   \caption{Spatial distribution of aggressive speeding observance} 
%   \label{fig:threshold_speeding_map}
% \end{figure}

% % Commenton spatial patterns if needed (i.e. UVA student concentration area? Income based areas?)

% \begin{figure}[!ht]
%   \centering
%   \includegraphics[width=0.8\textwidth]{extreme_speeding_osm.png}
%   \caption{Spatial distribution of reckless speeding observance} 
%   \label{fig:threshold_speeding_map}
% \end{figure}

\section{Discussion}
The analysis of speeding behavior across residential road segments using cumulative distribution functions underscores the highly skewed nature of both aggressive and reckless speeding percentage, with a small subset of roads accounting for a disproportionate share of violations. As shown in Figure~\ref{fig:cdf_wayid_aggressive} and Figure~\ref{fig:cdf_wayid_reckless}, 38\% of residential segments had at least one instance of aggressive speeding, and 20\% of segments showed at least one instance of reckless speeding. While most segments had very few violations, the top 2.7\% recorded aggressive speeding on at least 20\% of their trajectory points, and the top 1\% exhibited reckless speeding on more than 10\% of their records. These patterns highlight that while the majority of roads see minimal or no high-speed violations, a small group of residential segments experience repeated and severe speeding behaviors, which may warrant targeted interventions or enforcement measures.

% As shown in Figure \ref{fig:cdf_wayid_aggressive}, the 50th percentile is 0.0\%, the 85th percentile is 0.8\%, and 95\% of way IDs had aggressive speeding observance 9.3\%, indicating that more than 15\% of the segments having aggressive speeding observance, and more than 5\% of the segments having aggressive speeding observance more than 9\%. Similarly, Figure~\ref{fig:cdf_wayid_reckless} reveals that more than 5\% of segments had reckless speeding observance more than 0.9\%, with most segments falling well below this threshold. These distributions illustrate the disproportionate concentration of aggressive and reckless speeding behaviors on a smaller fraction of the residential network.

Tables~\ref{tab:top_wayid} and~\ref{tab:top_wayid_extreme} provide further detail, listing the ten residential segments with the highest aggressive and reckless speeding percentages. Notably, several segments exhibit exceptionally high violation rates—exceeding 90\% for aggressive speeding and over 60\% for reckless speeding. These segments also vary considerably in sample size and vehicle participation, with some segments showing more than 90\% violations from a small set of trajectory points, while others reflect widespread speeding issues among thousands of trajectory points. This variability has policy implications: low-volume but high-violation roads might warrant targeted signage or speed bumps, whereas higher-volume roads may require broader design or enforcement interventions. Importantly, several segments appear in both rankings, suggesting they pose compounded safety risks due to both frequent and extreme speeding events.

To ensure the validity of the most extreme cases, we conducted a manual check on the top 10 way IDs for both aggressive and reckless speeding percentage. We found that only 3 of the 11 unique way IDs appearing in the two top-10 way ID tables intersect with other residential roads, suggesting that these high speeding percentage values are likely valid. The remaining top segments often intersect with arterials with higher posted speed. In such cases, vehicles may not be able to decelerate quickly enough when transitioning into residential areas, which could explain the anomalously high speeding percentages.

These results reaffirm that while most residential roads have relatively low speeding prevalence, there exists a nontrivial tail of segments where speeding issues persistently happen. The fact that some segments include both moderate and extreme violators suggests the reason of speeding may be poorly marked posted speed limit signs on the streets, while others may be transitional links from higher-speed arterials where drivers fail to decelerate promptly. High-risk segments could benefit from engineering countermeasures such as speed humps or clearer posted speed limit signs. Understanding the physical or contextual characteristics of these segments is essential for targeted interventions.

Furthermore, Table \ref{tab:day_night_speeding_comparison} presents a comparison between daytime and nighttime aggressive speeding rates across way IDs with at least 100 total observations. For each segment, the daytime (or nighttime) speeding rate was calculated as the number of aggressive speeding records during that period divided by the total number of records for the same period. The results show that a substantial majority of segments—1,243 in total—exhibited more aggressive speeding during nighttime hours, while only 46 segments showed higher aggressive speeding during the day. 

\begin{table}[ht]
\centering
\caption{Comparison of Daytime and Nighttime Speeding Rates}
\begin{tabular}{|l|c|}
\hline

Number of Way IDs with higher daytime speeding rate & 46 \\
\hline
Number of Way IDs with higher nighttime speeding rate & 1,243 \\
\hline
\end{tabular}
\label{tab:day_night_speeding_comparison}
\end{table}

% \begin{figure}[!ht]
%   \centering
%   \includegraphics[width=0.8\textwidth]{NT_day_night_wayid.png}
%   \caption{Histogram of Way ID Day vs Night Aggressive Speeding Rate} 
%   \label{fig:day_night_wayid}
% \end{figure}

% Finally, upon checking on the OpenStreetMap, we observed that more serious speeding incidents tend to occur in suburban areas farther from the city center, suggesting that less urbanized road segments may experience higher speeding activity due to lower enforcement presence or more open roadway conditions.

% On the infrastructure side, a small subset of residential segments consistently shows high levels of aggressive and reckless speeding, with some segments exhibiting over 90\% violation rates. These patterns suggest that certain roadways may unintentionally facilitate speeding due to their design characteristics, lack of clear signage, or reduced enforcement presence—particularly in suburban areas, as highlighted in the spatial analysis.

% On the vehicle ID side, the vehicle-level analysis revealed that a limited group of drivers exhibit repeated high-speed violations across multiple segments. Some drivers showed aggressive speeding on more than half of their residential trajectory points, and in extreme cases, exceeded 90\% for both aggressive and reckless speeding observance. These findings suggest that high-risk driving behaviors are not random but are instead indicative of habitual speeding tendencies.

One possible explanation is that drivers may feel more comfortable speeding during nighttime hours due to lower perceived risk of enforcement, leading to more frequent aggressive speeding. Additionally, the emptier roads at night may encourage faster driving despite limited visibility. These patterns suggest that speeding behavior is not only spatially and behaviorally clustered, but also temporally concentrated during nighttime periods, highlighting the need for increased enforcement or targeted interventions during those hours. 

Lastly, integrating this telematics-based speeding analysis with crash data, traffic citation records, or land use information (e.g., school zones) would help validate identified hotspots and better quantify the safety consequences. Using the way ID-level perspectives, agencies can more precisely diagnose and mitigate speeding risks in residential networks—moving beyond generalized enforcement toward targeted, data-informed solutions.

\section{Conclusions}
% \subsection{Summary of Contributions}
% Here I will write about the core achievements of the project, linking back to the stated objectives.

% Delivered a full pipeline for processing, analyzing, and mapping residential speeding issuesusing Wejo data.

% Created interpretable outputs with actionable insights for transportation agencies.

% Demonstrated how telematics data can fill critical gaps in traditional traffic safety data.

This study presents a comprehensive data-driven framework for identifying and analyzing residential speeding behavior using high-resolution connected vehicle data. The methodology developed in this research enables the detection of both aggressive and reckless speeding behavior at a granular spatial scale, focusing on residential road segments within the study area.

A major contribution of this work is the creation of a scalable pipeline for processing large volumes of trajectory data, augmenting missing speed limit information, and generating interpretable summaries of speeding behavior at the way ID (road segment) level. The outputs include detailed metrics such as aggressive and reckless speeding percentages, average speeds, and speed standard deviations. The analysis reveals not only how frequently speeding occurs, but also how extreme it can be. Moreover, results show that speeding is more prevalent during nighttime hours. This suggests that drivers may be more prone to risky behavior when driving conditions feel more favorable despite limited visibility at nighttime.

Furthermore, this study demonstrates how connected vehicle telematics data can fill critical gaps in traditional traffic safety monitoring systems, which often rely on sparse infrastructure-based sensors or incomplete citation records. By leveraging telematics data, transportation agencies are empowered with a more comprehensive and continuous picture of driving behavior, particularly on residential streets where traditional data sources are limited.

In conclusion, this framework sets the foundation for future applications in traffic safety analytics, demonstrating the growing potential of telematics data to inform safer, more livable communities in Charlottesville and Albemarle County. By uncovering both the intensity and temporal patterns of speeding behavior—including extreme violations and the prevalence of nighttime speeding—this approach offers valuable insights for targeted enforcement and infrastructure planning.

% \subsection{Limitations}
% Here I will write about methodological constraints and data assumptions that may influence results.

% OSM speed limit data coverage may be incomplete or inaccurate.

% Wejo data only records newer vehicles. Not all speeding behavior is captured.
\subsection{Limitations}
While this analysis offers detailed insights, several limitations should be acknowledged. First, the accuracy of the speed limit data sourced from OSM may be incomplete or inconsistent, particularly for residential roads where speed limits are sometimes not explicitly tagged. This may result in either overestimation or underestimation of speeding rates on certain segments.

Second, the connected vehicle data we used, while rich in granularity, represents only a subset of the total vehicle population—primarily newer vehicles equipped with connected telematics systems. As such, the data may not fully capture the driving behavior of the entire population, particularly older vehicles or those not participating in the connected ecosystem.

Additionally, some extreme speed values are likely influenced by GPS signal errors or map-matching inaccuracies. While these occurrences were retained in the analysis to flag locations with potentially serious problems, they should be interpreted cautiously.

Despite these limitations, the results clearly demonstrate that telematics-based analyses can reveal critical insights into residential speeding behavior, offering a valuable complement to traditional enforcement and planning tools.

\subsection{Future Work}
% Merge with crash data, land use, or socio-demographics.

% We tried to link with crash data but number of crash reported is limited.

% We tried to link with AADT data but most residential roads do not have AADT information.

While this study offers valuable insights, there remain several opportunities for future research to extend this work. A key direction is to integrate speeding behavior data with other datasets such as socio-demographic information or school zone information. This would enable a more holistic understanding of the relationship between speeding behavior and traffic safety outcomes.

Initial attempts to merge the speeding dataset with local crash data faced limitations due to the relatively small number of reported crashes on residential streets within the study area. Similarly, efforts to link with Annual Average Daily Traffic (AADT) data were constrained by the fact that most residential roads lack comprehensive AADT coverage. Future work could benefit from access to more granular AADT estimates or synthetic traffic volume models to better contextualize exposure risk. Additionally, obtaining crash data from insurance claims or third-party aggregators may offer a richer dataset for linking observed speeding behavior with actual safety outcomes, especially in cases where police-reported crash records are sparse or incomplete.

Lastly, refining the methodological approach to better handle GPS noise, improve map-matching accuracy, and apply advanced statistical techniques could further improve the robustness and reliability of the results.

\section{Acknowledgments}
This research was funded by the SMARTER University Transportation Center and the Virginia Department of Transportation (as the UTC cost share). The Virginia Department of Transportation provided data access to the connected vehicle data.

ChatGPT was used to correct grammar, refine the tone to be more academically formal, and assist with debugging Python and LaTeX code.

\section{Author contribution statement}
The authors confirm contribution to the paper as follows: study conception and design: B. Brian Park, Andrew Mondschein; data collection: Shi Feng; analysis and interpretation of results: Shi Feng; draft manuscript preparation: Shi Feng, B. Brian Park, Andrew Mondschein. All authors reviewed the results and approved the final version of the manuscript.

\section{Declaration of Conflicting Interests}
The authors declared no potential conflicts of interest with respect to the research, authorship, and/or publication of this article.

\newpage

\bibliographystyle{unsrt}  
\bibliography{references}

\end{document}